\documentclass[letterpaper,10pt]{article} 

\usepackage{opticameet3} 

\newcommand\authormark[1]{\textsuperscript{#1}}

\usepackage{amsmath,amssymb}
\usepackage{algorithm}
\usepackage{algpseudocode}
\usepackage{booktabs}
\usepackage{url}
\usepackage[colorlinks=true,bookmarks=false,citecolor=blue,urlcolor=blue]{hyperref} 

\begin{document}

\title{The geodesic dispersion phenomenon in\\random fields dynamics}

\author{Alexandre L. M. Levada\authormark{1}}

\address{Computing Department, Federal University of S\~ao Carlos, S\~ao Carlos, SP, Brazil}


\email{alexandre.levada@ufscar.br} 

\begin{abstract}
Random fields are ubiquitous mathematical structures in physics, with applications ranging from thermodynamics and statistical physics to quantum field theory and cosmology. Recent works on information geometry of Gaussian random fields proposed mathematical expressions for the components of the metric tensor of the underlying parametric space, allowing the computation of the curvature in each point of the manifold. In this study, our hypothesis is that time irreversibility in Gaussian random fields dynamics is a direct consequence of intrinsic geometric properties (curvature) of their parametric space. In order to validate this hypothesis, we compute the components of the metric tensor and derive the twenty seven Christoffel symbols of the metric to define the Euler-Lagrange equations, a system of partial differential equations that are used to build geodesic curves in Riemannian manifolds. After that, by the application of the fourth-order Runge-Kutta method and Markov Chain Monte Carlo simulation, we numerically build geodesic curves starting from an arbitrary initial point in the manifold. The obtained results show that, when the system undergoes phase transitions, the geodesic curve obtained by time reversing the computational simulation diverges from the original curve, showing a strange effect that we called the geodesic dispersion phenomenon, which suggests that time irreversibility in random fields is related to the intrinsic geometry of their parametric space.
\end{abstract}

\section{Introduction}

The exploration of the fundamental principles governing the behavior of physical systems has been a longstanding pursuit in several branches of theoretical and applied physics \cite{Quantum1}. In this pursuit, the incorporation of randomness and disorder has emerged as a pivotal avenue for understanding the intricate dynamics exhibited by a diverse array of systems \cite{Disorder1,Disorder2}. Among the mathematical frameworks employed to model and analyze such inherent uncertainties, random fields have proven to be indispensable \cite{Uncertainty}. Examples of scenarios where random fields have profound implications in physics are statistical physics \cite{StatsPhys}, condensed matter physics \cite{CondensedMatter}, phase transitions \cite{PhaseT}, critical phenomena \cite{CriticalT}, quantum systems \cite{quantum}, non-equilibrium statistical mechanics \cite{StatMecha}, cosmology \cite{Cosmo} and astrophysics \cite{Astro}.

Information geometry is a branch of mathematics that explores the geometric structures associated with statistical models and probability distributions, by modeling the parametric space of random variables as a Riemannian manifold \cite{Amari2021,Nielsen1,Nielsen2}. It provides a powerful framework for analyzing the relationships and structures within these manifolds, offering deep insights into the intrinsic geometry of random fields \cite{Levada1}. In particular, information geometry provides a formal way to study random fields dynamics, offering insights into local geometric properties of the parametric space as the system ehxibits critical phenomena and phase transitions \cite{Levada2}.

The concept of irreversible dynamics in random fields refers to the phenomenon where the evolution of the field over time exhibits a lack of symmetry or reversibility, meaning that the system does not return to its initial state when allowed to evolve backward \cite{Irreversible1}. Irreversible dynamics often relate to an increase in entropy over time. Entropy is a measure of disorder or randomness in a system. As a random field evolves, the system tends to explore a wider range of configurations, leading to increased entropy. The tendency for entropy to increase is a key aspect of irreversibility in many physical systems \cite{Irreversible2}. In the context of thermodynamics, irreversible dynamics can be associated with processes that involve dissipation, such as heat transfer, friction, or irreversible chemical reactions. These processes contribute to the irreversibility of the overall system, and the random field may reflect these irreversible thermodynamic behaviors. Phase transitions in random fields, such as those occurring in condensed matter systems, can contribute to irreversible dynamics. The system may undergo abrupt changes in its state, and once the system has transitioned to a new phase, it might not easily revert to its previous state \cite{Irreversible3}.

Recent studies about information geomerty have proposed expressions for the metric tensor of the Gaussian random fields manifold, which allows the computation of the curvatures of the space along random field dynamics \cite{Levada1}. Computational simulations revealed the curvature effect, which means that a highly asymmetric curvature patterns are observed in the underlying parametric space when there are significant increase/decrease in the inverse temperature parameter, leading to phase transitions. When entropy is increasing, the curvature transitions from negative to positive in a smoother way than in comparison to the transition from positive to negative when entropy is decreasing \cite{Levada2}.

Motivated by this findings, this paper proposes the following hypothesis: time irreversibility in Gaussian random fields dynamics is a direct consequence of the intrinsic geometry of their parametric space. To investigate this hypothesis, our study presents a computational method to build geodesic curves in the manifold of Gaussian random fields, by numerically solving the Euler-Lagrange equations that govern how a point moves along geodesics in this parametric space, given an initial location and an initial tangent vector (velocity). Basically, the main contributions of this paper are: 1) to derive the Christoffel symbols of the metric tensor of the Gaussian random fields manifold \cite{Weinberg,Wheeler,ChristoffelSymbols}; 2) to build the Euler-Lagrange equations, which defines a system of nonlinear differential equations that govern the dynamics of a moving reference along geodesic curves in the underlying manifold \cite{EulerLagrange1,EulerLagrange2,EulerLagrange3}; and 3) to solve the Euler Lagrange equations using the fourth-order Runge-Kutta \cite{Runge,Kutta,RK4} method and Markov-Chain Monte Carlo simulation \cite{Metropolis,Hastings,MCMC2021}. Computational experiments reveal an interesting behavior, that we called the geodesic dispersion phenomenon. In several cases, the geodesic curve obtained by time reversing the simulations diverge from the original geodesic curve, which suggests that, when curvature decreases from zero to negative values, the emergence of a hyperbolic-like geometry makes a given geodesic curve to deviate from its original trajectory.

The remaining of the paper is organized as follows: in Section 2, we introduce the random field model and in Section 3, a brief overview of Markov Chain Monte Carlo simulations is presented. In Section 4, we show the components of the metric tensor for the random fields manifold and in Section 5, we discuss geodesic distances in Riemannian manifolds. In Section 6, we briefly describe the fourth-order Runge-Kutta (RK4) method and in Section 7, we derive the Christoffel symbols of the metric. Section 8 shows the proposed iterative MCMC based RK4 method for generating goedesic curves and in Section 9, we present the computational experiments and results. Finally, Section 10 presents our conclusions and final remarks.

\section{Gaussian random fields}

Gaussian random fields (GRFs) are stochastic processes that exhibit Gaussian (normal) distribution at every spatial location. These fields are characterized by the property that any finite collection of random variables, corresponding to different locations in the field, follows a multivariate normal distribution. Gaussian random fields are extensively used in various scientific disciplines, including physics, due to their mathematical tractability and versatility in modeling complex spatial structures and uncertainties.

\begin{figure}[ht]
	\begin{center}
	\includegraphics[scale=0.4]{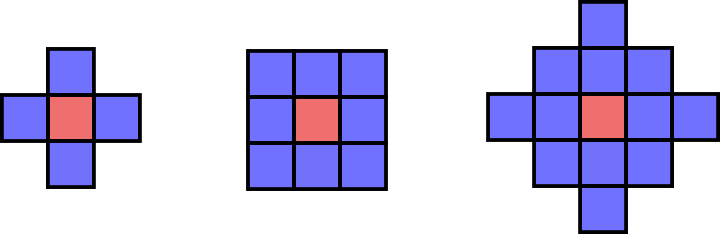}
	\end{center}
	\caption{First, second and third order neighborhood systems on a 2D lattice.}
	\label{fig:neigh}
\end{figure}

GRFs are models in which the set of possible states for a given variable is continuous and innumerable. In practice, each variable in the field can assume an infinite number of states \cite{GMRFBook}. We assume some simplifying hypothesis due to mathematical tractability. First, we model only binary interactions in a 2D lattice, constraining the analysis to pairwise interaction random field models. Additionally, the inverse temperature parameter, which controls the spatial dependence structure, is invariant and isotropic along the random field. Finally, by invoking the Hammersley-Clifford theorem \cite{Hammersley}, which states the equivalence between Gibbs random fields (global models) and Markov random fields (local models), we characterize our model by the set of local conditional density functions:

\begin{equation}
	p\left( x_{i} | \eta_{i}, \vec{\theta} \right) = \frac{1}{\sqrt{2\pi\sigma^2}}exp\left\{-\frac{1}{2\sigma^{2}} \left[ x_{i} - \mu - \beta \sum_{j \in \eta_{i}} \left( x_{j} - \mu \right) \right]^{2} \right\}
	\label{eq:GMRF}
\end{equation} where $\eta_i$ denotes the the second-order neighborhood system, comprised by the 8 nearest neighbors of $x_i$ in the 2D lattice, $\vec{\theta} = (\mu, \sigma^{2}, \beta)$ denotes the vector of parameters, with $\mu$ and $\sigma^{2}$ being, the expected value (mean) and the variance of the random variables in the lattice, and $\beta$ the inverse temperature. Note that if $\beta = 0$, the model degenerates to a regular Gaussian density and the variables in the field become independent. Figure \ref{fig:neigh} shows the first, second and third order neighborhood systems defined on a 2D lattice. In summary, Gaussian random fields find applications in various branches of physics due to their ability to model spatial variability and uncertainty in physical processes.

\section{Markov Chain Monte Carlo simulation}

Markov Chain Monte Carlo (MCMC) algorithms are a class of computational methods used for sampling from complex probability distributions, such as random field models. These algorithms are particularly valuable when direct sampling from the distribution is difficult or impractical. The fundamental idea behind MCMC is to construct a Markov chain that has the desired probability distribution as its equilibrium distribution. Once the Markov chain reaches its equilibrium, samples from it can be considered as samples from the target distribution \cite{MCMC1,MCMC2,MCMC3}.

A Markov chain is a mathematical model that describes a sequence of events, where the probability of transitioning from one state to another depends only on the current state, not on the sequence of events that preceded it. In the context of MCMC, the states represent different points in the parameter space of the distribution being sampled. Monte Carlo methods involve the use of random sampling to obtain numerical results. In the case of MCMC, Monte Carlo methods are employed to generate a sequence of samples from the target distribution \cite{MCMC4,MCMC5,MCMC6}.

MCMC algorithms construct a Markov chain that satisfies the detailed balance condition, ensuring that the desired distribution is its equilibrium distribution. This involves defining a transition probability from one state to another in such a way that the resulting stationary distribution is the target distribution. One of the most commonly used MCMC algorithms is the Metropolis-Hastings algorithm \cite{Metropolis}. In each iteration of this algorithm, a proposed new state is generated, and a decision is made to accept or reject this proposal based on a acceptance probability. The acceptance probability is designed to satisfy the detailed balance condition, ensuring convergence to the target distribution. Algorithm \ref{alg:mh} shows the pseudocode for the Metropolis-Hastings method, used in all experiments along this paper.

\begin{algorithm}
\caption{Metropolis-Hastings algorithm for Markov Chain Monte Carlo simulation}\label{alg:mh}
\begin{algorithmic}[1]
\Function{Metropolis}{$n$, $m$, $N$, $\mu$, $\sigma^2$, $\beta$}
\State Generate a $n \times m$ matrix $X$ with samples from a Gaussian density with mean $\mu$ and variance $\sigma^2$.
\While{$i \leq N$}
	\For{$i=1; i \leq n; i++$}
		\For{$i=j; j \leq m; j++$}
			\State $y_{ij} = Normal(\mu, \sigma^2)$
			\State Let $\displaystyle P = min \left\{ 1, \frac{p(y_{ij} | \eta_{i}, \vec{\theta})}{p(x_{ij} | \eta_{i}, \vec{\theta})} \right\}$
			\State Replace $x_{ij}$ by $y_{ij}$ with probability $p$
		\EndFor
	\EndFor
\EndWhile
\EndFunction
\end{algorithmic}
\end{algorithm}

Recent works in applied probability and computational statistics link known results about the diffusion limits of Markov chain Monte Carlo (MCMC) algorithms to the concept of algorithm complexity in computer science. It has been shown that, under appropriate assumptions, the random-walk in Metropolis-adjusted algorithms takes $O(d^{1/3})$ iterations to converge to stationarity in $d$ dimensions \cite{MCMC7}.

\section{The metric tensor of the Gaussian random fields manifold}

It has been shown by information geometry that the Fisher information matrix is the metric tensor that equips the underlying parametric space of a statistical manifold \cite{Amari,Nielsen}. In practical terms, the metric tensor makes it possible to express the square of an infinitesimal displacement in the manifold, $ds^2$, as a function of an infinitesimal displacement in the tangent space. In order to define the Euler-Lagrange equations and compute geodesic distances in the Gaussian random fields manifold, we need to compute the metric tensor of the underlying parametric space. It has been shown in previous works that the metric tensor of the Gaussian random fields manifold has the following shape \cite{Levada1,Levada2}:

\begin{equation}
	g(\vec{\theta}) = \left( \begin{array}{ccc}
	g_{11} & g_{12} & g_{13} \\ 
	g_{21} & g_{22} & g_{23} \\ 
	g_{31} & g_{32} & g_{33}
	\end{array} \right)
\end{equation} where $g_{ij}$ is the Fisher information regarding parameters $\theta_i$ and $\theta_j$:

\begin{equation}
	g_{ij} = E\left[ \left(\frac{\partial}{\partial\theta_i} log~p(x; \vec{\theta}) \right)\left(\frac{\partial}{\partial\theta_j} log~p(x; \vec{\theta}) \right) \right]
\end{equation} in which $p(x; \vec{\theta})$ denotes the local conditional density function of the random field model. The closed-form expressions for all components of the metric tensor are listed below \cite{Levada1,Levada2}:

\begin{align}
	g_{11} &= \frac{\left(1 - \beta\Delta \right)^2}{\sigma^2} \left[ 1 - \frac{1}{\sigma^2}\left(  2\beta\sum_{j\in\eta_i}\sigma_{ij} - \beta^2\sum_{j\in\eta_i}\sum_{k\in\eta_i}\sigma_{jk} \right) \right] \\ 
	g_{12} &= g_{13} = g_{21} = 0 \\
	g_{22} & = \frac{1}{2\sigma^4} - \frac{1}{\sigma^6}\left[ 2\beta\sum_{j\in\eta_i}\sigma_{ij} - \beta^2 \sum_{j\in\eta_i}\sum_{k\in\eta_i}\sigma_{jk} \right] + \frac{1}{\sigma^8}\left[ 3\beta^2 \sum_{j\in\eta_i}\sum_{k\in\eta_i}\sigma_{ij}\sigma_{ik} - \beta^3 \sum_{j\in\eta_i}\sum_{k\in\eta_i}\sum_{l\in\eta_i}\left( \sigma_{ij}\sigma_{kl} + \sigma_{ik}\sigma_{jl} + \sigma_{il}\sigma_{jk} \right) \nonumber \right. \\  & \hspace{1cm} \left. + \beta^4 \sum_{j\in\eta_i}\sum_{k\in\eta_i}\sum_{l\in\eta_i}\sum_{m\in\eta_i}\left( \sigma_{jk} \sigma_{lm} + \sigma_{jl}\sigma_{km} + \sigma_{jm}\sigma_{kl} \right)  \right] \\
	g_{23} & = g_{32} = \frac{1}{\sigma^4}\left[ \sum_{j\in\eta_i}\sigma_{ij} - \beta\sum_{j\in\eta_i}\sum_{k\in\eta_i}\sigma_{jk} \right] \\ \nonumber & - \frac{1}{2\sigma^6}\left[ 6\beta\sum_{j\in\eta_i}\sum_{k\in\eta_i}\sigma_{ij}\sigma_{ik} - 3 \beta^2 \sum_{j\in\eta_i}\sum_{k\in\eta_i}\sum_{l\in\eta_i}\left( \sigma_{ij}\sigma_{kl} + \sigma_{ik}\sigma_{jl} + \sigma_{il}\sigma_{jk} \right) \right. \\ \nonumber & \hspace{3cm} \left. + \beta^3 \sum_{j\in\eta_i}\sum_{k\in\eta_i}\sum_{l\in\eta_i}\sum_{m\in\eta_i} \left( \sigma_{jk}\sigma_{lm} + \sigma_{jl}\sigma_{km} + \sigma_{jm}\sigma_{kl} \right) \right] 
\end{align}

\begin{align}
	g_{31} & = 0 \\
	g_{33} & = \frac{1}{\sigma^2}\sum_{j\in\eta_i} \sum_{k\in\eta_i} \sigma_{jk} + \frac{1}{\sigma^4} \left[ 2 \sum_{j\in\eta_i} \sum_{k\in\eta_i} \sigma_{ij} \sigma_{ik}  - 2\beta \sum_{j\in\eta_i} \sum_{k\in\eta_i} \sum_{l\in\eta_i} \left( \sigma_{ij}\sigma_{kl} + \sigma_{ik}\sigma_{jl} + \sigma_{il}\sigma_{jk} \right) \right. \\ \nonumber & \left. + \beta^2 \sum_{j\in\eta_i} \sum_{k\in\eta_i} \sum_{l\in\eta_i} \sum_{m\in\eta_i} \left( \sigma_{jk}\sigma_{lm} + \sigma_{jl}\sigma_{km} + \sigma_{jm}\sigma_{kl} \right) \right]
\end{align} where $\Delta$ is the cardinality of the neighborhood system ($\Delta = 8$ in a second-order system), $\sigma_{ij}$ is the covariance between the central variable $x_i$ and one of its neighbors $x_j \in \eta_i$ and $\sigma_{jk}$ is the covariance between two variables $x_j$ and $x_k$ belonging to the neighborhood $\eta_i$. 

In order to reduce the computational cost in the numerical simulations, it is possible to express the components of the metric tensor using Kronecker products (tensor products). First, note that we can convert each $3 \times 3$ neighborhood patch formed by $x_{i} \cup \eta_{i}$ into a vector $\vec{p}_i$ of 9 elements by piling its rows. Then, we compute the covariance matrix of these vectors, for $i = 1, 2,..., n$ denoted by $\Sigma_{p}$. From this covariance matrix, we extract two main components: 1) a vector of size 8, $\vec{\rho}$, composed by the the elements of the central row of $\Sigma_{p}$, excluding the middle one, which denotes the variance of $x_i$ (we want only the covariances between $x_i$ and $x_j$, for $j \neq i$; and 2) a sub-matrix of dimensions $8 \times 8$, $\Sigma_{p}^{-}$, obtained by removing the central row and central column from $\Sigma_{p}$ (we want only the covariances between $x_j \in \eta_i$ and $x_k \in \eta_i$). Figure \ref{fig:cov_matrix} shows the decomposition of the covariance matrix $\Sigma_{p}$ into the sub-matrix $\Sigma_{p}^{-}$ and the vector $\vec{\rho}$. By employing Kronecker products, we rewrite the first fundamental form (metric tensor) in a tensorial notation, providing a computationally efficient way to compute the elements of $g(\vec{\theta})$:

\begin{figure}[ht]
\begin{center}
\includegraphics[scale=0.4]{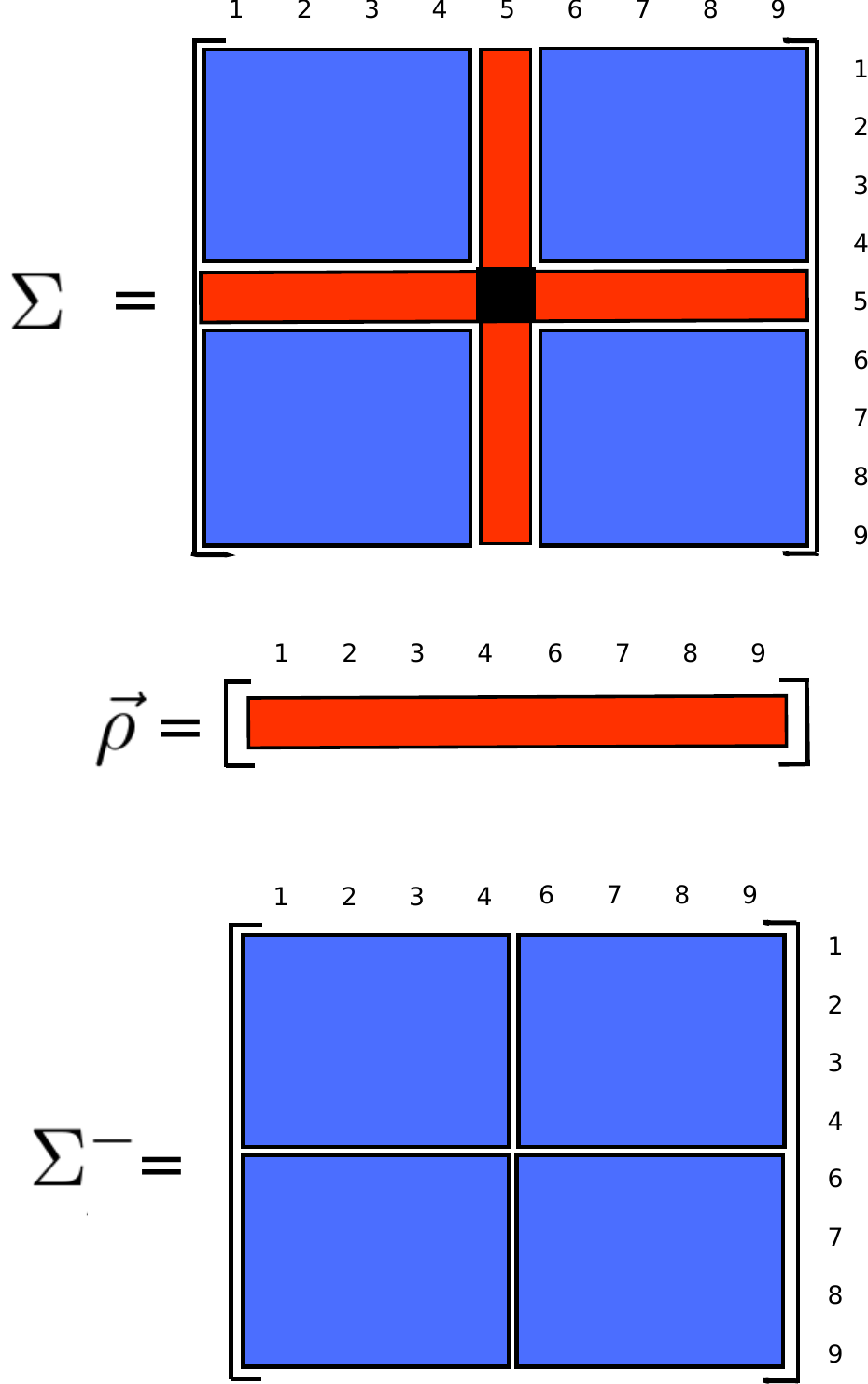}
\end{center}
\caption{Decomposition of $\Sigma_{p}$ into $\Sigma_{p}^{-}$ and $\vec{\rho}$ on a second-order neighborhood system ($\Delta=8$). By rewriting the components of the metric tensor in terms of Kronocker products, we can make numerical simulations faster.
}
\label{fig:cov_matrix}
\end{figure}

\begin{align}
	g_{11} & = \frac{1}{\sigma^2}\left(1-\beta \Delta \right)^2\left[ 1 - \frac{1}{\sigma^2}\left( 2\beta \left\| \vec{\rho} \right\|_{+} - \beta^2 \left\| \Sigma_{p}^{-} \right\|_{+} \right) \right] \label{eq:g11} \\
	\label{eq:g22}
	g_{22} & = \frac{1}{2\sigma^4} - \frac{1}{\sigma^6} \left[ 2\beta\left\| \vec{\rho} \right\|_{+} - \beta^2 \left\| \Sigma_{p}^{-} \right\|_{+} \right] + \frac{1}{\sigma^8}\left[ 3\beta^2 \left\| \vec{\rho} \otimes \vec{\rho} \right\|_{+} - 3 \beta^3 \left\| \vec{\rho} \otimes \Sigma_{p}^{-} \right\|_{+} + 3\beta^4 \left\| \Sigma_{p}^{-} \otimes \Sigma_{p}^{-} \right\|_{+}  \right] \\
	\label{eq:g23}
	g_{23} & = g_{32} = \frac{1}{\sigma^4}\left[ \left\| \vec{\rho} \right\|_{+} - \beta \left\| \Sigma_{p}^{-} \right\|_{+} \right] - \frac{1}{2\sigma^6} \left[ 6\beta \left\| \vec{\rho} \otimes \vec{\rho} \right\|_{+} - 9 \beta^2 \left\| \vec{\rho} \otimes \Sigma_{p}^{-} \right\|_{+} + 3\beta^3 \left\| \Sigma_{p}^{-} \otimes \Sigma_{p}^{-} \right\|_{+}  \right] \\ 
	\label{eq:g33}
	g_{33} & = \frac{1}{\sigma^2} \left\| \Sigma_{p}^{-} \right\|_{+} + \frac{1}{\sigma^4} \left[ 2 \left\| \vec{\rho} \otimes \vec{\rho} \right\|_{+} - 6 \beta \left\| \vec{\rho} \otimes \Sigma_{p}^{-} \right\|_{+} + 3\beta^2 \left\| \Sigma_{p}^{-} \otimes \Sigma_{p}^{-} \right\|_{+}  \right] 
\end{align} where $\left\| A \right\|_{+}$ represents the summation of all the entries of the vector/matrix $A$ and $\otimes$ denotes the Kronecker (tensor) product. The advantage of this tensorial notation using Kronecker products is that we avoid the computation of triple and quadruple summations, which have complexity $O(n^3)$ and $O(n^4)$, respectively.

\subsection{Entropy in Gaussian random fields}

The concept of entropy is crucial in many areas of science. In summary, entropy is a thermodynamic quantity that measures the degree of molecular freedom of a system, being associated with its number of configurations (or microstates), that is, how many ways particles can be distributed in quantized energy levels. It is also generally associated to the randomness, dispersion of matter and energy, and ``disorder'' of a complex system. In random fields, it is possible to compute the entropy for each possible observed configuration. The entropy of a Gaussian random field can be computed by the expected value of self-information, which leads to:

\begin{align}
	\label{eq:entropia1}
	 H_{\beta}(\vec{\theta}) = - E\left[ log~p\left(x_{i}| \eta_{i}, \vec{\theta} \right) \right] = H_{G}(\vec{\theta}) - \frac{\beta}{\sigma^2}\sum_{j \in \eta_i}\sigma_{ij} + \frac{\beta^2}{2\sigma^2} \sum_{j \in \eta_i}\sum_{k \in \eta_i}\sigma_{jk}
\end{align} where $H_G(\vec{\theta})$ denotes the entropy of a Gaussian random variable with mean $\mu$ and variance $\sigma^2$. Note that the entropy is a quadratic function of the inverse temperature parameter $\beta$. Besides, for $\beta=0$, we have $H_{\beta}(\vec{\theta}) = H_{G}(\vec{\theta})$, as expected. Using the tensorial notation, the entropy can be expressed as: 

\begin{align}
	\label{eq:entropia2}
	 H_{\beta}(\vec{\theta}) = H_{G}(\vec{\theta}) - \frac{1}{\sigma^2} \left( \beta\left\| \vec{\rho} \right\|_{+} - \frac{\beta^2}{2} \left\| \Sigma_{p}^{-} \right\|_{+} \right)
\end{align}

Note that $H_{G}(\vec{\theta})$ does not depend on the inverse temperature parameter, so differetiating $H_{\beta}(\vec{\theta})$ with respect to $\beta$ leads to:

\begin{align}
	\frac{\partial}{\partial\beta}H_{\beta} & = -\frac{1}{\sigma^2} \left( \left\| \vec{\rho} \right\|_{+} - \beta \left\| \Sigma_{p}^{-} \right\|_{+} \right)
\end{align} which means that the second derivative of entropy with respect to $\beta$ is:

\begin{equation}
\frac{\partial^2}{\partial\beta^2}H_{\beta} = \frac{1}{\sigma^2} \left\| \Sigma_{p}^{-} \right\|_{+}
\end{equation}

\section{Geodesic distances in Riemannian manifolds}

In a Riemannian manifold $M$ equipped with a metric tensor $g$, the length of a curve $\gamma(t)$, defined for $t \in [a, b]$ is:

\begin{equation}
	L(\gamma) = \int_a^b \sqrt{g(\gamma'(t), \gamma'(t))} dt
\end{equation} where $\gamma'(t)$ denotes the first derivative (tangent vector). It is easier to work with the energy of the curve. The energy of a curve $\gamma(t)$ is given by:

\begin{equation}
	E(\gamma) = \frac{1}{2} \int_a^b \sqrt{g(\gamma'(t), \gamma'(t))} dt
\end{equation} 

It is not difficult to show that if $M$ is a manifold equipped with a metric tensor $g(\theta)$, then:

\begin{equation}
	L(\gamma) \leq 2 (b - a) E(\gamma)
\end{equation} with equality if and only if $\gamma$ is parametrized by arc length, that is, if we move along the curve with constant velocity (the tangent vectors are unitary). In summary, a curve $\gamma(t)$ has minimum energy if and only if it has minimum length and arc length parametrization. A geodesic is a curve $\gamma(t)$ that minimizes the functional $L(\gamma)$. In order to obtain an exact solution, we need to solve the following optimization problem \cite{Riemann}:

\begin{equation}
	\nabla_{\gamma'} \gamma' = 0
\end{equation} where $\nabla$ is the covariant derivative. The geodesics are precisely the ``straight lines'' that have constant velocity. From differential geometry, it has been shown that a solution to this equation must satisfy the geodesic equations, expressed by \cite{Tristan}:

\begin{equation}
	\frac{d^2 \theta_k}{dt^2} + \sum_{ij} \Gamma_{ij}^k \frac{d \theta_i}{dt} \frac{d \theta_j}{dt} = 0
\end{equation} where, in our pairwise isotropic Gaussian-Markov random field, $\theta_1 = \mu$, $\theta_2 = \sigma^2$ and $\theta_3 = \beta$ are the model parameters and $\Gamma_{ij}^k$ are the Christoffel symbols of the metric, defined as \cite{Manfredo}:

\begin{equation}
	\Gamma_{ij}^k = \frac{1}{2}\sum_{m=1}^{3} \left( \frac{\partial}{\partial\theta_i}g_{jm} + \frac{\partial}{\partial\theta_j}g_{im} - \frac{\partial}{\partial\theta_m}g_{ij} \right) g^{mk}
\end{equation} where $g_{ij}$ a component of the metric tensor (Fisher information matrix) and $g^{ij}$ is a component of the inverse metric tensor. As the parametric space of pairwise isotropic Gaussian-Markov random fields are 3D manifolds, we have a total of $3 \times 3 \times 3 = 27$ Christoffel symbols arranged in three $3 \times 3$ matrices $\Gamma^1$, $\Gamma^2$ and $\Gamma^3$, leading to the following system of non-linear differential equations:

\begin{align}
\theta_1''(t) - \sum_{ij} \Gamma_{ij}^{1} \theta_i'(t) \theta_j'(t) = 0 \\
\theta_2''(t) - \sum_{ij} \Gamma_{ij}^{2} \theta_i'(t) \theta_j'(t) = 0 \\
\theta_3''(t) - \sum_{ij} \Gamma_{ij}^{3} \theta_i'(t) \theta_j'(t) = 0
\end{align} where $\theta_i'(t)$ and $\theta_i''(t)$ denotes the first and second derivatives of the coordinates, the values of $\theta_1^{(0)} = \mu_0$, $\theta_2^{(0)} = \sigma_0^2$ and $\theta_3^{(0)} = \beta_0$ represent the starting point A, and the values of $\theta_1'^{(0)}$, $\theta_2'^{(0)}$ and $\theta_3'^{(0)}$ defines the initial tangent vector at A (initial velocity vector). The fourth-order Runge-Kutta method can be applied to obtain a numerical solution for this system of differential equations, where the final value of the parameter $t$ (number of iterations) determines how long the geodesic curve will grow. Note that there are no explicit values for the final parameters $\theta_1^{(t)} = \mu_t$, $\theta_2^{(t)} = \sigma_t^2$ and $\theta_3^{(t)} = \beta_t$, since the manifold is a curved space and the geodesic is build locally, using one tangent space at a time.

\section{The fourth-order Runge-Kutta method}

The Runge-Kutta method is a numerical technique used for solving ordinary differential equations (ODEs). Its importance lies in its ability to provide accurate and efficient solutions to a wide range of differential equations encountered in various scientific and engineering applications \cite{RK1,RK2}. The main advantages of the Runge-Kutta method over other approaches are \cite{RK3,RKHist,RK5}:

\begin{itemize}
	\item \textbf{Accuracy:} by employing multiple stages in its computations, such as the classic fourth-order Runge-Kutta (RK4), it achieves a balance between accuracy and computational efficiency.
	\item \textbf{Applicability to Systems of ODEs:} many physical problems involve systems of interconnected ODEs. The Runge-Kutta method naturally extends to handle systems of differential equations, making it a versatile tool for modeling complex dynamic systems. 
	\item \textbf{Adaptive Step Size:} RK4 is essential when dealing with ODEs where the solution undergoes rapid changes in certain regions and remains relatively constant in others. 
	\item \textbf{Numerical Stability and Robustness:} RK4 exhibits good numerical stability characteristics. This property is crucial for preventing numerical instabilities that may arise during the solution of differential equations. 
\end{itemize}

In summary, the Runge-Kutta method is a fundamental numerical technique with widespread applications in scientific research, especially in physics. Its accuracy, stability, and versatility make it an indispensable tool for solving systems of ordinary differential equations and gaining insights into the behavior of random fields dynamics. The most popular version of Runge–Kutta method is the fourth-order algorithm, often denoted by RK4, the classic Runge–Kutta method, which is described in the following. Consider a first-order ordinary differential equation in the form:

\begin{equation}
	\frac{dy}{dt} = f(t, y)
\end{equation} where $y$ is the unknown function of $t$ and $f(t, y)$ is a given function representing the rate of change of $y$ with respect to $t$. At the initial time $t_0$, the corresponding $y$ value is $y_0$. It should be mentioned that the function $f$ and the initial conditions $t_0, y_0$ must be given as input to the problem. The fourth-order Runge-Kutta algorithm is given by the following steps \cite{RK4}:

\begin{enumerate}
	\item Define $n = 0$ and the step size $h > 0$.
	\item Compute the four increments:
	\begin{align}
		k_1 & = h f(t_n, y_n) \\
		k_2 & = h f\left( t_n + \frac{h}{2}, y_n + \frac{k_1}{2} \right) \\
		k_3 & = h f\left( t_n + \frac{h}{2}, y_n + \frac{k_2}{2} \right) \\
		k_4 & = h f\left( t_n + h, y_n + k_3 \right)
	\end{align}	 
	\item Update the value of the function:
	\begin{align}
		y_{n+1} & = y_n + \frac{1}{6}\left( k_1 + 2k_2 + 2k_3 + k_4 \right) \\
		t_{n+1} & = t_n + h
	\end{align}
	\item Repeat steps 2 and 3 for $n = 1, 2, 3, ..., N$, where $N$ is the number of iterations.
\end{enumerate}

Basically, the fourth-order Runge-Kutta algorithm works by iteratively updating the solution of an ordinary differential equation based on weighted averages of incremental steps, providing an accurate and efficient numerical approximation of the solution over a specified range.

\section{The Christoffel symbols of the metric}

In order to compute the Christoffel symbols of the underlying parametric space of pairwise isotropic Gaussian-Markov random field models, we need to find the components of the inverse metric tensor $g^{-1}(\theta)$:

\begin{equation}
	g^{-1}(\vec{\theta}) = \left( \begin{array}{ccc}
	g^{11} & 0 & 0 \\ 
	0 & g^{22} & g^{23} \\ 
	0 & g^{32} & g^{33}
	\end{array} \right)
\end{equation} 

It is straightforward to check that the components of the inverse metric tensor are:

\begin{align}
	g^{11} = \frac{1}{g_{11}} \qquad g^{22} = \frac{g_{33}}{g_{22}g_{33} - g_{23}^2}  \qquad g^{23} = \frac{g_{23}}{g_{23}^2 - g_{22}g_{33}} \qquad g^{32} = \frac{g_{32}}{g_{32}^2 - g_{22}g_{33}} \qquad g^{33} = \frac{g_{22}}{g_{22}g_{33} - g_{32}^2}
\end{align} where $g^{23} = g^{32}$. It is worth mentioning that in the computational experiments, we have to regularize the metric tensor to avoid numerical instability in the matrix inversion process by adding a small value $\lambda = 0.001$ to its main diagonal. The next step consists in the derivation of the components of the Christoffel symbols. First, note that:

\begin{equation}
	\frac{\partial}{\partial\theta_1} g_{ij} = \frac{\partial}{\partial\mu} g_{ij} = 0, \qquad\qquad \forall i, j
\end{equation} since none of the components of the metric tensor is a function of $\theta_1 = \mu$. Moving to the derivatives with respect to $\theta_2 = \sigma^2$, we have:

\begin{align}
	\frac{\partial}{\partial\theta_2}g_{11} & = -\frac{1}{\sigma^4}(1 - \beta\Delta)^2 + \frac{2}{\sigma^6}(1 - \beta\Delta)^2 \left( 2\beta \left\| \vec{\rho} \right\|_{+} - \beta^2 \left\| \Sigma_{p}^{-} \right\|_{+} \right) \\
	\frac{\partial}{\partial\theta_2}g_{22} & = -\frac{1}{\sigma^6} + \frac{3}{\sigma^8} \left( 2\beta \left\| \vec{\rho} \right\|_{+} - \beta^2 \left\| \Sigma_{p}^{-} \right\|_{+} \right) - \frac{4}{\sigma^{10}} \left( 3\beta^2 \left\| \vec{\rho} \otimes \vec{\rho} \right\|_{+} - 3 \beta^3 \left\| \vec{\rho} \otimes \Sigma_{p}^{-} \right\|_{+} + 3 \beta^4  \left\| \Sigma_{p}^{-} \otimes \Sigma_{p}^{-} \right\|_{+} \right) \\
	\frac{\partial}{\partial\theta_2}g_{23} & = \frac{\partial}{\partial\theta_2}g_{32} = -\frac{2}{\sigma^6}\left( \left\| \vec{\rho} \right\|_{+} - \beta \left\| \Sigma_{p}^{-} \right\|_{+} \right) + \frac{3}{2\sigma^8}\left( 6\beta \left\| \vec{\rho} \otimes \vec{\rho} \right\|_{+} - 9\beta^2 \left\| \vec{\rho} \otimes \Sigma_{p}^{-} \right\|_{+} + 3 \beta^3 \left\| \Sigma_{p}^{-} \otimes \Sigma_{p}^{-} \right\|_{+} \right) \\
	\frac{\partial}{\partial\theta_2}g_{33} & = -\frac{1}{\sigma^4} \left\| \Sigma_{p}^{-} \right\|_{+} - \frac{2}{\sigma^6}\left( 2 \left\| \vec{\rho} \otimes \vec{\rho} \right\|_{+} - 6\beta \left\| \vec{\rho} \otimes \Sigma_{p}^{-} \right\|_{+} + 3 \beta^2 \left\| \Sigma_{p}^{-} \otimes \Sigma_{p}^{-} \right\|_{+} \right)
\end{align}

Finally, we have to differentiate the components of the metric tensor with respect to the inverse temperature parameter, that is, $\theta_3 = \beta$:

\begin{align}
	\frac{\partial}{\partial\theta_3}g_{11} & = -\frac{1}{\sigma^2}2\Delta(1 - \beta\Delta) \left[ 1 - \frac{1}{\sigma^2}\left( 2\beta\left\| \vec{\rho} \right\|_{+} - \beta^2 \left\| \Sigma_{p}^{-} \right\|_{+} \right) \right] - \frac{1}{\sigma^4} (1 - \beta\Delta)^2 \left( 2\left\| \vec{\rho} \right\|_{+} - 2\beta\left\| \Sigma_{p}^{-} \right\|_{+} \right) \\
	\frac{\partial}{\partial\theta_3}g_{22} & = -\frac{1}{\sigma^6}\left( 2\left\| \vec{\rho} \right\|_{+} - 2\beta \left\| \Sigma_{p}^{-} \right\|_{+} \right) + \frac{1}{\sigma^8} \left( 6\beta \left\| \vec{\rho} \otimes \vec{\rho} \right\|_{+} - 9 \beta^2 \left\| \vec{\rho} \otimes \Sigma_{p}^{-} \right\|_{+} + 12 \beta^3  \left\| \Sigma_{p}^{-} \otimes \Sigma_{p}^{-} \right\|_{+} \right) \\	
	\frac{\partial}{\partial\theta_3}g_{23} & = \frac{\partial}{\partial\theta_3}g_{32} = -\frac{1}{\sigma^4}\left\| \Sigma_{p}^{-} \right\|_{+} - \frac{1}{2\sigma^6}\left( 6\left\| \vec{\rho} \otimes \vec{\rho} \right\|_{+} - 18\beta \left\| \vec{\rho} \otimes \Sigma_{p}^{-} \right\|_{+} + 9 \beta^2 \left\| \Sigma_{p}^{-} \otimes \Sigma_{p}^{-} \right\|_{+} \right) \\
	\frac{\partial}{\partial\theta_3}g_{33} & = -\frac{1}{\sigma^4} \left( 6\beta \left\| \vec{\rho} \otimes \Sigma_{p}^{-} \right\|_{+} - 6 \beta \left\| \Sigma_{p}^{-} \otimes \Sigma_{p}^{-} \right\|_{+} \right)
\end{align}

Given the above, it is possible to compute the 27 Christoffel symbols, denoted by: $\Gamma_{11}^1, \Gamma_{11}^2, \Gamma_{11}^3, ..., \Gamma_{33}^1, \Gamma_{33}^2, \Gamma_{33}^3$, organized in a tensor composed by three $3 \times 3$ matrices $\Gamma^1, \Gamma^2, \Gamma^3$. However, it can be shown that the Christoffel symbols are symmetric with respect to the lower indices, that is, $\Gamma_{ij}^k = \Gamma_{ji}^k$, resulting in a total of 18 different symbols. First, note that the components $\Gamma_{11}^1, \Gamma_{11}^2, \Gamma_{11}^3$ are:


\begin{align}
	\Gamma_{11}^1 & = 0 \label{eq:A} \\
	\Gamma_{11}^2 & = -\frac{1}{2}\left( \frac{\partial g_{11}}{\partial\theta_2}g^{22} + \frac{\partial g_{11}}{\partial\theta_3}g^{32} \right) \\
	\Gamma_{11}^3 & = -\frac{1}{2}\left( \frac{\partial g_{11}}{\partial\theta_2}g^{23} + \frac{\partial g_{11}}{\partial\theta_3}g^{33} \right)
\end{align}

Next, the components $\Gamma_{12}^1, \Gamma_{12}^2, \Gamma_{12}^3$ can be expressed by:



\begin{align}
	\Gamma_{12}^1 = \frac{1}{2} \frac{\partial g_{11}}{\partial\theta_2}g^{11} = \Gamma_{21}^1 \qquad
	\Gamma_{12}^2 = \Gamma_{21}^2 = 0 \qquad
	\Gamma_{12}^3 = \Gamma_{21}^3 = 0
\end{align}

Similarly, the components $\Gamma_{13}^1, \Gamma_{13}^2, \Gamma_{13}^3$ also have the same structure:

\begin{align}
	\Gamma_{13}^1 = \frac{1}{2} \frac{\partial g_{11}}{\partial\theta_3}g^{11} = \Gamma_{31}^1 \qquad
	\Gamma_{13}^2 = \Gamma_{31}^2 = 0 \qquad
	\Gamma_{13}^3 = \Gamma_{31}^3 = 0
\end{align}

Moving forward to the components $\Gamma_{22}^1, \Gamma_{22}^2, \Gamma_{22}^3$, we have:


\begin{align}
	\Gamma_{22}^1 & = 0 \\
	\Gamma_{22}^2 & = \frac{1}{2} \left[ \frac{\partial g_{22}}{\partial\theta_2}g^{22} + \left( 2 \frac{\partial g_{23}}{\partial\theta_2} - \frac{\partial g_{22}}{\partial\theta_3} \right)g^{32} \right] \\
	\Gamma_{22}^3 & = \frac{1}{2} \left[ \frac{\partial g_{22}}{\partial\theta_2}g^{23} + \left( 2 \frac{\partial g_{23}}{\partial\theta_2} - \frac{\partial g_{22}}{\partial\theta_3} \right)g^{33} \right]
\end{align}

The components $\Gamma_{23}^1, \Gamma_{23}^2, \Gamma_{23}^3$ are given by:


\begin{align}
	\Gamma_{23}^1 & = \Gamma_{32}^1 = 0 \\
	\Gamma_{23}^2 & = \frac{1}{2}\left( \frac{\partial g_{22}}{\partial\theta_3}g^{22} + \frac{\partial g_{33}}{\partial\theta_2}g^{32} \right) = \Gamma_{32}^2 \\
	\Gamma_{23}^3 & = \frac{1}{2}\left( \frac{\partial g_{22}}{\partial\theta_3}g^{23} + \frac{\partial g_{33}}{\partial\theta_2}g^{33} \right) = \Gamma_{32}^3
\end{align}

Finally, the components $\Gamma_{33}^1, \Gamma_{33}^2, \Gamma_{33}^3$ can be computed by:


\begin{align}
	\Gamma_{33}^1 & = 0 \\
	\Gamma_{33}^2 & = \frac{1}{2} \left[ \left( 2 \frac{\partial g_{32}}{\partial\theta_3} - \frac{\partial g_{33}}{\partial\theta_2} \right) g^{22} + \frac{\partial g_{33}}{\partial\theta_3}g^{32} \right] \\
	\Gamma_{33}^3 & = \frac{1}{2} \left[ \left( 2 \frac{\partial g_{32}}{\partial\theta_3} - \frac{\partial g_{33}}{\partial\theta_2} \right) g^{23} + \frac{\partial g_{33}}{\partial\theta_3}g^{33} \right] \label{eq:B}
\end{align}

Therefore, from the 27 Christoffel symbols, 13 are zero and the remaining 14 non-zero symbols assume 10 different values due to symmetry properties.

\subsection{Approximating the curvature}

In differential geometry, to compute the curvature of a surface, it is necessary to obtain the expressions of the second fundamental form and of the differential of the Gauss map in a coordinate system. In computational terms, this can be done through the definition of the shape operator \cite{Manfredo}. Let $S$ be a surface with first fundamental form $\mathbb{I}$ and second fundamental form $\mathbb{II}$. Then, the shape operator $P$ is:

\begin{equation}
	P = -(\mathbb{II})(\mathbb{I})^{-1} = - \begin{bmatrix} 
	L & M \\
	M & N \\
	\end{bmatrix} \begin{bmatrix} 
	E & F \\
	F & G \\
	\end{bmatrix}^{-1}
\end{equation}	

The shape operator encodes relevant information about the curvature of surfaces, being a powerful mathematical tool for geometric analysis. From the shape operator, we can obtain the Gaussian, mean and principal curvatures as the determinant, the trace and the eigenvalues of the shape operator. In this study, we compute the shape operator of the Gaussian random fields manifold to estimate the curvatures. Note that the metric tensor is the generalization of the first fundamental form, so we compute the second fundamental form as the second-order Fisher information matrix:

\begin{equation}
	\left\{ \mathbb{II}(\vec{\theta}) \right\}_{ij} = -\mathbb{E}\left[ \frac{\partial^2}{\partial\theta_i \partial\theta_j} log~p(X; \vec{\theta}) \right], \text{~~~~ for } i,j=1,\ldots,n
\end{equation}

It has been shown that, in pairwise isotropic Gaussian random fields, the elements of the second fundamental form are given by \cite{Levada1,Levada2}:

\begin{align}
	\mathbb{II}_{\mu\mu}(\vec{\theta}) & = \frac{1}{\sigma^2}\left(1-\beta \Delta \right)^2 \\
	\mathbb{II}_{\sigma^2\sigma^2}(\vec{\theta}) & = \frac{1}{2\sigma^4} - \frac{1}{\sigma^6} \left[ 2\beta\left\| \vec{\rho} \right\|_{+} - \beta^2 \left\| \Sigma_{p}^{-} \right\|_{+} \right] = \frac{1}{2\sigma^4} + \frac{1}{\sigma^4} \left[ 2(H_{\beta} - H_G) \right] \\
	\mathbb{II}_{\sigma^2\beta}(\vec{\theta}) & = II_{\beta\sigma^2}(\vec{\theta}) = \frac{1}{\sigma^4}\left[ \left\| \vec{\rho} \right\|_{+} - \beta \left\| \Sigma_{p}^{-} \right\|_{+} \right] = -\frac{1}{\sigma^2}\frac{\partial}{\partial\beta}H_{\beta} \\
	\mathbb{II}_{\beta\beta}(\vec{\theta}) & = \frac{1}{\sigma^2} \left\| \Sigma_{p}^{-} \right\|_{+} = \frac{\partial^2}{\partial\beta^2}H_{\beta}	
\end{align} showing that they are closely related to the system's entropy and its first and second derivatives.

\section{The proposed MCMC based RK4 simulation algorithm}

In order to numerically solve our system of non-linear second-order differential equations, first we have to convert it to a first-order system. This can be done by a simple variable substitution. Let $\gamma_1(t) = \theta_1(t)$ and $\alpha_1(t) = \theta_1'(t)$. Then, we have:

\begin{align}
	\gamma_1'(t) & = \theta_1'(t) = \alpha_1(t) \\
	\alpha_1'(t) & = \theta_1''(t) = - \sum_{ij} \Gamma_{i,j}^1 \alpha_i(t) \alpha_j(t)
\end{align}

By direct application of this procedure to the other equations, we finally reach our system of non-linear first-order differential equations:

\begin{align}
	\gamma_1'(t) & = \alpha_1(t) \\
	\gamma_2'(t) & = \alpha_2(t) \\
	\gamma_3'(t) & = \alpha_3(t) \\
	\alpha_1'(t) & = - \sum_{ij} \Gamma_{i,j}^1 \alpha_i(t) \alpha_j(t)\\
	\alpha_2'(t) & = - \sum_{ij} \Gamma_{i,j}^2 \alpha_i(t) \alpha_j(t)\\
	\alpha_3'(t) & = - \sum_{ij} \Gamma_{i,j}^3 \alpha_i(t) \alpha_j(t)
\end{align}

Note that we can rewrite our system in a more compact way as:

\begin{align}
	\gamma_1'(t) & = F(t, \vec{\alpha}) = \alpha_1 \\
	\gamma_2'(t) & = G(t, \vec{\alpha}) = \alpha_2 \\
	\gamma_3'(t) & = H(t, \vec{\alpha}) = \alpha_3 \\
	\alpha_1'(t) & = P(t, \vec{\alpha}, \Gamma_1) = - \vec{\alpha}^T \Gamma^1 \vec{\alpha} \\
	\alpha_2'(t) & = Q(t, \vec{\alpha}, \Gamma_2) = - \vec{\alpha}^T \Gamma^2 \vec{\alpha} \\
	\alpha_3'(t) & = R(t, \vec{\alpha}, \Gamma_3) = - \vec{\alpha}^T \Gamma^3 \vec{\alpha}
\end{align} where $\vec{\alpha}^T = [ \alpha_1, \alpha_2, \alpha_3 ]$ is the tangent vector and, for $k = 1, 2, 3$:

\begin{equation}
	\Gamma^k = \left( \begin{array}{ccc}
	\Gamma_{11}^k & \Gamma_{12}^k & \Gamma_{13}^k \\ 
	\Gamma_{21}^k & \Gamma_{22}^k & \Gamma_{23}^k \\ 
	\Gamma_{31}^k & \Gamma_{32}^k & \Gamma_{33}^k \\ 
	\end{array} \right)
\end{equation}

\begin{algorithm}[]
\caption{Computing geodesic curves in random fields manifold with RK4 and MCMC simulation}\label{alg:gd}
\begin{algorithmic}[1]
\Function{GeodesicDistanceRandomField}{$a$, $b$, $n$, $N$, $\vec{\gamma}^{(0)}$, $\vec{\alpha}^{(0)}$}
\State $h = (b - a)/n$
\State $t = a$
\State $L = 0$					\Comment{L denotes the geodesic distance}
\For{$i = 0$, $i < N$, $i++$} 
	\State Generate an outcome of the GMRF model using $\vec{\gamma}^{(i)}$.
	\State Compute the metric tensor $g(\vec{\gamma}^{(i)})$ with equations (\ref{eq:g11}), (\ref{eq:g22}), (\ref{eq:g23}) and (\ref{eq:g33}).
	\State Regulize the metric tensor adding a small $\epsilon$ in its main diagonal. 
	\State Compute the inverse metric tensor $g^{-1}(\vec{\gamma}^{(i)})$.
	\State Compute the Christoffel symbols ($\Gamma_1$, $\Gamma_2$, $\Gamma_3$) with equations (\ref{eq:A}) to (\ref{eq:B}).
	\State // Perform a 4th order Runge Kutta iteration
	\State $k_0 = hF(t, \vec{\alpha})$
	\State $l_0 = hG(t, \vec{\alpha})$
	\State $m_0 = hH(t, \vec{\alpha})$			
	\State $x_0 = hP(t, \vec{\alpha})$
	\State $y_0 = hQ(t, \vec{\alpha})$
	\State $z_0 = hR(t, \vec{\alpha})$	
	\State $\vec{\delta} = [k_0, l_0, m_0]$
	\State $k_1 = hF(t+0.5h, \vec{\alpha}+0.5\vec{\delta})$
	\State $l_1 = hG(t+0.5h, \vec{\alpha}+0.5\vec{\delta})$
	\State $m_1 = hH(t+0.5h, \vec{\alpha}+0.5\vec{\delta})$			
	\State $x_1 = hP(t+0.5h, \vec{\alpha}+0.5\vec{\delta}, \Gamma_1)$
	\State $y_1 = hQ(t+0.5h, \vec{\alpha}+0.5\vec{\delta}, \Gamma_2)$
	\State $z_1 = hR(t+0.5h, \vec{\alpha}+0.5\vec{\delta}, \Gamma_3)$
	\State $\vec{\delta} = [k_1, l_1, m_1]$
	\State $k_2 = hF(t+0.5h, \vec{\alpha}+0.5\vec{\delta})$
	\State $l_2 = hG(t+0.5h, \vec{\alpha}+0.5\vec{\delta})$
	\State $m_2 = hH(t+0.5h, \vec{\alpha}+0.5\vec{\delta})$			
	\State $x_2 = hP(t+0.5h, \vec{\alpha}+0.5\vec{\delta}, \Gamma_1)$
	\State $y_2 = hQ(t+0.5h, \vec{\alpha}+0.5\vec{\delta}, \Gamma_2)$
	\State $z_2 = hR(t+0.5h, \vec{\alpha}+0.5\vec{\delta}, \Gamma_3)$
	\State $\vec{\delta} = [k_2, l_2, m_2]$
	\State $k_3 = hF(t+h, \vec{\alpha}+\vec{\delta})$
	\State $l_3 = hG(t+h, \vec{\alpha}+\vec{\delta})$
	\State $m_3 = hH(t+h, \vec{\alpha}+\vec{\delta})$			
	\State $x_3 = hP(t+h, \vec{\alpha}+\vec{\delta}, \Gamma_1)$
	\State $y_3 = hQ(t+h, \vec{\alpha}+\vec{\delta}, \Gamma_2)$
	\State $z_3 = hR(t+h, \vec{\alpha}+\vec{\delta}, \Gamma_3)$
	\State Update the vectors $\vec{\gamma}^{(i)}$ (position) and $\vec{\alpha}^{(i)}$ (velocity) as
	\State $\gamma_1^{(i+1)} = \gamma_1^{(i)} + (1/6)(k_0 + 2k_1 + 2k_2 + k_3)$
	\State $\gamma_2^{(i+1)} = \gamma_2^{(i)} + (1/6)(l_0 + 2l_1 + 2l_2 + l_3)$
	\State $\gamma_3^{(i+1)} = \gamma_3^{(i)} + (1/6)(m_0 + 2m_1 + 2m_2 + m_3)$
	\State $\alpha_1^{(i+1)} = \alpha_1^{(i)} + (1/6)(x_0 + 2x_1 + 2x_2 + x_3)$
	\State $\alpha_2^{(i+1)} = \alpha_2^{(i)} + (1/6)(y_0 + 2y_1 + 2y_2 + y_3)$
	\State $\alpha_3^{(i+1)} = \alpha_3^{(i)} + (1/6)(z_0 + 2z_1 + 2z_2 + z_3)$
	\State Update the geodesic distance: $L = L + \lVert \alpha(t) \rVert h$
\EndFor
\EndFunction
\end{algorithmic}
\end{algorithm}


The basic idea is that the initial condition $\vec{\gamma}^{(0)} = [ \gamma_1^{(0)}, \gamma_2^{(0)}, \gamma_3^{(0)} ]$ defines the starting point in the manifold and the initial condition $\vec{\alpha}^{(0)} = [ \alpha_1^{(0)}, \alpha_2^{(0)}, \alpha_3^{(0)} ]$ defines the initial tangent vector (velocity). The arc length of the resulting geodesic curve defines our geodesic distance. The arc length between $t = r$ and $t = s$ can be approximated by:

\begin{equation}
	s(t) \approx \sum_{t=r}^{s} \sqrt{ \left( x(t+\Delta t) - x(t) \right)^2 + \left( y(t+\Delta t) - y(t) \right)^2 + \left( z(t+\Delta t) - z(t) \right)^2 }
\end{equation}

Note that  we can rewrite $s(t)$ as:

\begin{equation}
	s(t) \approx \sum_{t=r}^{s} \sqrt{ \left( \frac{x(t+\Delta t) - x(t)}{\Delta t} \right)^2 + \left( \frac{y(t+\Delta t) - y(t)}{\Delta t} \right)^2 + \left( \frac{z(t+\Delta t) - z(t)}{\Delta t} \right)^2 } \Delta t
\end{equation}

From the definition of the derivative of a function, we have:

\begin{equation}
	f'(t) = \lim\limits_{\Delta t \to 0} \frac{f(t+\Delta t) - f(t)}{\Delta t}
\end{equation}

Thus, in the limiting case, when $\Delta t \to 0$, the approximation becomes exact:

\begin{align}
	s(t) = \int_{a}^{b} \sqrt{ x'(t)^2 + y'(t)^2 + z'(t)^2 } dt = \int_{a}^{b} \sqrt{ \vec{\gamma}'(t)^T \vec{\gamma}'(t)  } dt = \int_{a}^{b} \lVert \vec{\gamma}'(t) \rVert dt
\end{align} where the integrand $\lVert \vec{\gamma}'(t) \rVert$ is the norm of the tangent vector at $t$. Therefore, we compute the geodesic distance as:


\begin{equation}
	L(\gamma) = \int_{a}^{b}  \lVert \gamma'(t) \rVert dt \approx \sum_{i=a}^{b} \lVert \gamma'(i) \rVert   h
\end{equation} using a small enough step size $h=(b-a)/n$. 

\section{Computational experiments}

In the first set of experiments, we use the fouth-order Runge-Kutta method to build geodesic curves in the regular 2D Gaussian random variables manifold, which is known to have a constante negative curvature equal to minus one. Information geometry in Gaussian random variables has been extensively studied in the literature \cite{FisherGaussian,ChristGaussian}. The Gaussian random field model considered in this study degenerates to the regular Gaussian distribution by setting $\beta = 0$. In the numerical simulations, it is enough to initialize the vectors as $\vec{\gamma}^{(0)} = [\gamma_{1}^{(0)}, \gamma_{2}^{(0)}, 0]$ and $\vec{\alpha}^{(0)} = [\alpha_{1}^{(0)}, \alpha_{2}^{(0)}, 0]$. It is not difficult to see that the metric tensor of the Gaussian random variables submanifold is:

\begin{equation}
	g(\theta) = \left( \begin{array}{ccc}
	\displaystyle \frac{1}{\sigma^2} & \displaystyle 0 & \displaystyle 0 \\ 
	\displaystyle 0 & \displaystyle \frac{1}{2\sigma^4} & \displaystyle 0 \\
	\displaystyle 0 & \displaystyle 0 & \displaystyle 0 
	\end{array} \right)
\end{equation}

Moreover, the derivation of the Christoffel symbols of the metric reveals that:

\begin{equation}
	\Gamma^1 = \left( \begin{array}{ccc}
	\displaystyle 0 & \displaystyle -\frac{1}{2\sigma^2} & \displaystyle 0 \\ 
	\displaystyle -\frac{1}{2\sigma^2} & \displaystyle 0 & \displaystyle 0 \\
	\displaystyle 0 & \displaystyle & \displaystyle 0
	\end{array} \right) \qquad
	\Gamma^2 = \left( \begin{array}{ccc}
	\displaystyle 1 & \displaystyle 0 & \displaystyle 0 \\ 
	\displaystyle 0 & \displaystyle -\frac{1}{\sigma^2} & \displaystyle 0 \\
	\displaystyle 0 & \displaystyle 0 & \displaystyle0
	\end{array} \right) \qquad
	\Gamma^3 = \left( \begin{array}{ccc}
	\displaystyle 0 & \displaystyle 0 & \displaystyle 0 \\ 
	\displaystyle 0 & \displaystyle 0 & \displaystyle 0 \\
	\displaystyle 0 & \displaystyle 0 & \displaystyle0
	\end{array} \right)
\end{equation}

%

Given the above, we define our system of equations as:

\begin{align}
	\gamma_1'(t) & = F(t, \vec{\alpha}) = \alpha_1 \\
	\gamma_2'(t) & = G(t, \vec{\alpha}) = \alpha_2 \\
	\gamma_3'(t) & = H(t, \vec{\alpha}) = 0 \\
	\alpha_1'(t) & = P(t, \vec{\alpha}, \Gamma_1) = - \vec{\alpha}^T \Gamma^1 \vec{\alpha} \\
	\alpha_2'(t) & = Q(t, \vec{\alpha}, \Gamma_2) = - \vec{\alpha}^T \Gamma^2 \vec{\alpha} \\
	\alpha_3'(t) & = R(t, \vec{\alpha}, \Gamma_3) = - \vec{\alpha}^T \Gamma^3 \vec{\alpha} = 0
\end{align} where $\vec{\alpha}^T = [ \alpha_1, \alpha_2, 0 ]$ is the tangent vector. The main difference in comparison to the random field model is that we do not need to  perform MCMC simulations to generate outcomes of the random variable, which results in a more efficient algorithm. Table \ref{tab:ds} shows some obtained results using the proposed simulation method. Note that, as expected, the geodesic distances are greater than the simple improper 2D Euclidean distance between the initial and final parameter vectors in $R^2$. Additionally, the difference between the original geodesic distance and the geodesic distances obtained by time reversing the Runge-Kutta simulations are negligible, probably due to small numerical instabilities and minor rounding errors. It is worth to mention that the Gaussian random variables manifold has constant negative curvature, which means a hyperbolic-like geometry. The parameters for all the simulations were set to $a = 0$, $b = 10$ and $n = 1000$, which leads to step size $h = 0.01$.  

\begin{table}[ht]
\centering
\small
\caption{Results of the fourth-order Runge-Kutta simulation for the regular Gaussian density manifold.}
\begin{tabular}{ccccccccccc}
\toprule
\multicolumn{2}{c}{\textbf{Init. point (A)}} & \multicolumn{2}{c}{\textbf{Initial} $\vec{\alpha}$} & \multicolumn{2}{c}{\textbf{Final point (B)}} & \multicolumn{2}{c}{\textbf{Final $\vec{\alpha}$}} & \multicolumn{3}{c}{\textbf{Distances}} \\
\midrule
$\mu$         & $\sigma^2$         & $\alpha_1$       & $\alpha_2$       & $\mu$  & $\sigma^2$        & $\alpha_1$      & $\alpha_2$      & \textbf{Geo (AB)} & \textbf{Geo (BA)} & \textbf{Eucl.} \\
\midrule
0                & 1                    & 0.1000                & 0.0500                & 1.0765               & 0.9576                   & 0.0957               & -0.0554              & 1.1246                 & 1.1250                 & 1.0769             \\
2                & 1                    & 0.2000                & 0.05                  & 3.4723               & 0.2818                   & 0.0561               & -0.0689              & 1.7293                 & 1.7251                 & 1.6373             \\
1                & 2                    & 0.2500                & 0.2500                & 3.6214               & 1.1789                   & 0.1471               & -0.2396              & 3.3177                 & 3.3178                 & 2.7461             \\
10               & 5                    & 0.5000                & 0.5000                & 13.8957              & 1.2913                   & 0.1286               & -0.3741              & 6.4719                 & 6.4656                 & 5.3787             \\
10               & 10                   & 0.5000                & 2.0000                & 17.6233              & 11.3884                  & 0.5696               & -2.0745              & 17.0900                & 17.0826                & 7.7444           \\
100               & 100                   & 1.0000                & 1.0000                & 109.0051              & 68.3735                  & 0.6830               & -5.4847              & 34.8442                & 35.0553                & 32.8869           \\
\bottomrule
\end{tabular}
\label{tab:ds}
\end{table}

\begin{figure}[ht]
\begin{center}
\includegraphics[scale=0.12]{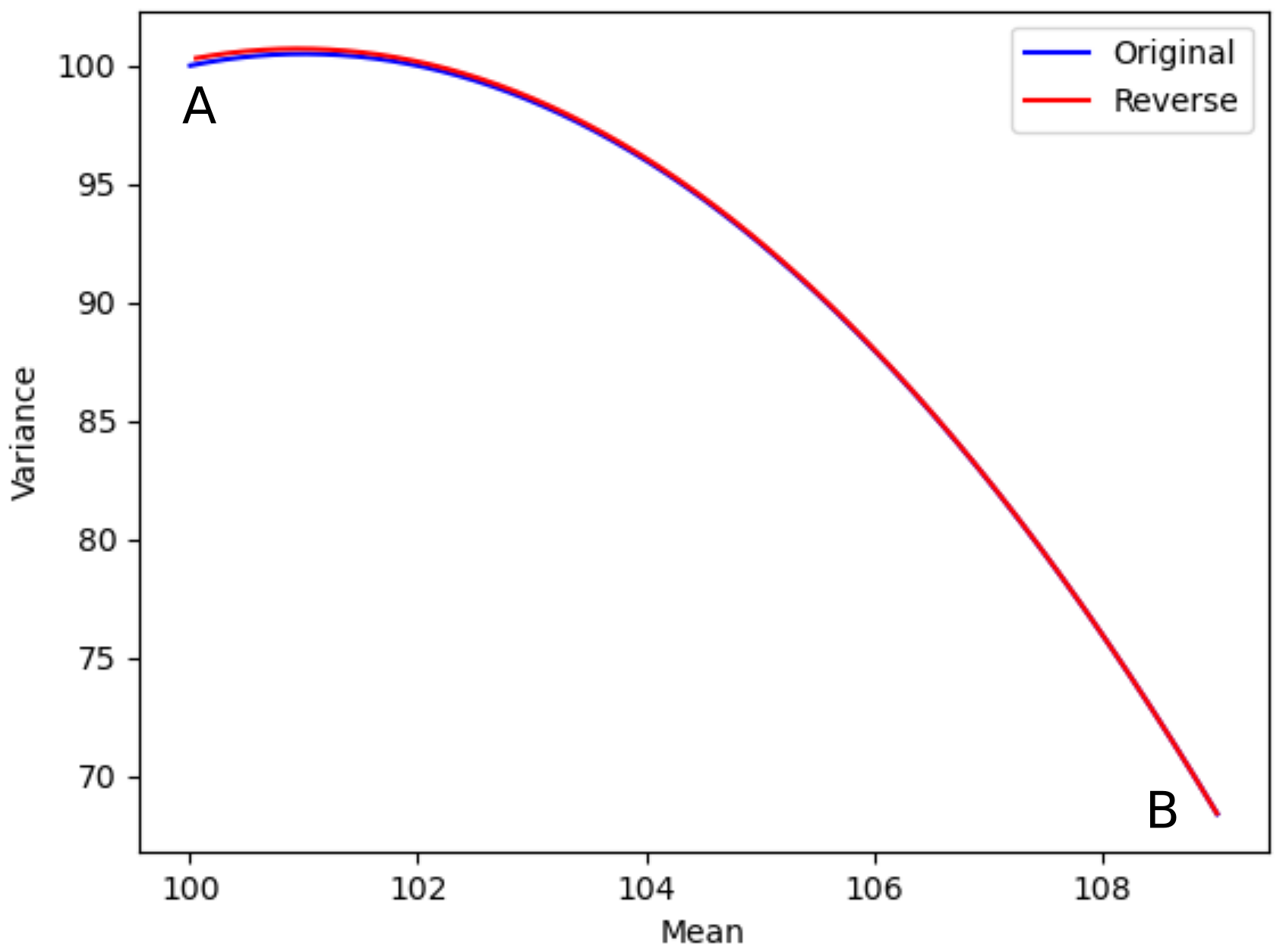}
\includegraphics[scale=0.12]{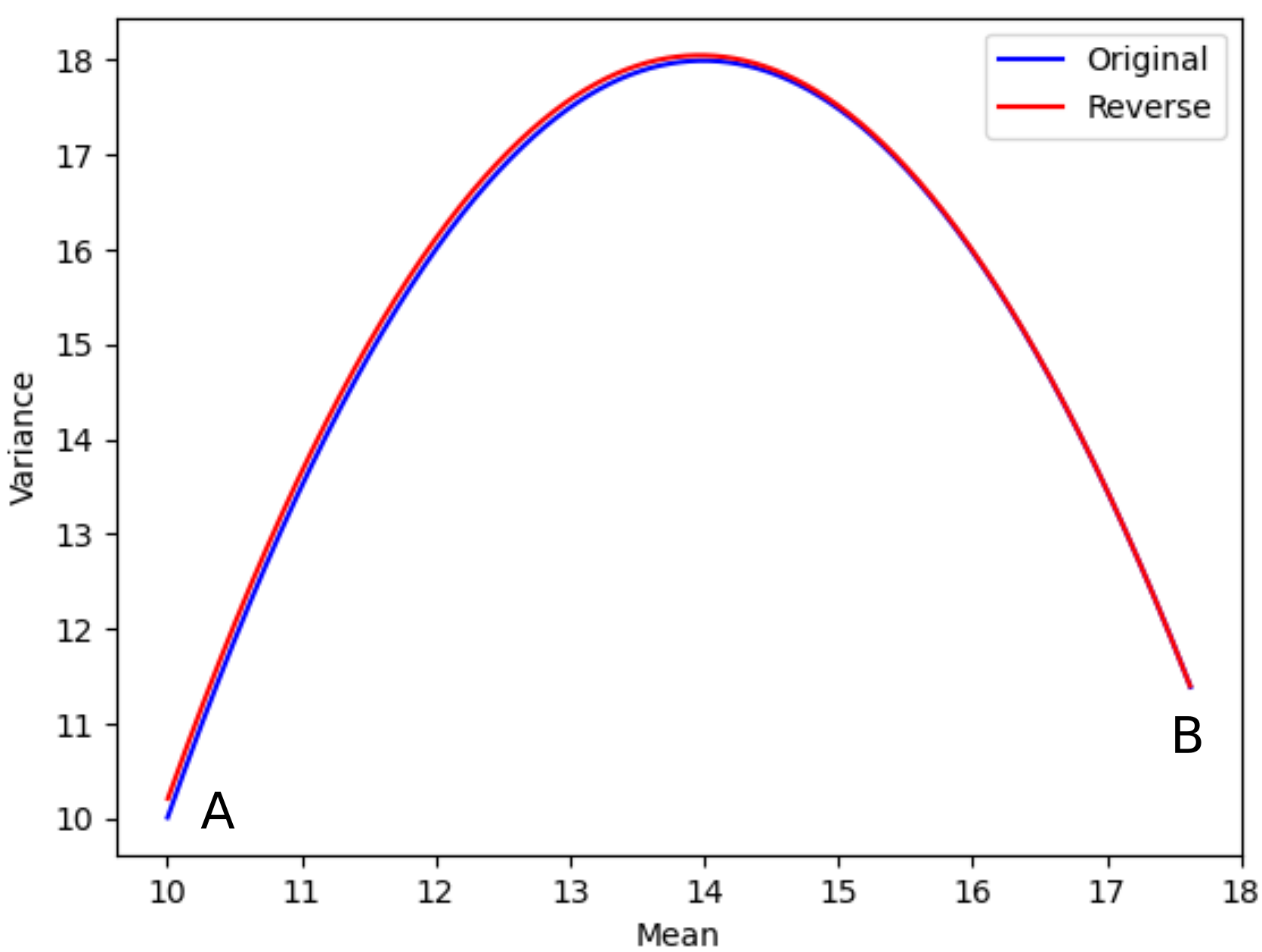}
\caption{Geodesic curves in the Gaussian density manifold obtained by fourth-order Runge-Kutta and its time reverse simulations. A denotes the initial point and B denotes the final point in the regular simulation and vice-versa in the time reversed version.}
\label{fig:gauss}
\end{center}
\end{figure}

Figure \ref{fig:gauss} illustrates some of the geodesic curves obtained during our simulations. The blue curve is built during the regular execution of the fourth-order Runge-Kutta method and the red curve is built during the time reversed version of the simulation. Note that the curves are almost identical, suggesting that the dynamics modeled by the system of differential equations can be considered reversible in time.  

In the second set of experiments, we use the proposed iterative MCMC based fouth-order Runge-Kutta method to build geodesic curves in the Gaussian random fields manifold using the metric tensor and the Christoffel symbols derived in the previous sections. In all simulations, due to the computational cost, the dimensions of each outcome of the random field are $128 \times 128$, the number of iterations in the MCMC simulation is $100$, $a = 0$, $b = 10$ and $n = 1000$, which leads to a step size $h = 0.01$. 

Roughly speaking, the intuition behind the numerical simulation of the Euler-Lagrange equations is that the initial condition $\vec{\gamma}^{(0)} = [ \gamma_1^{(0)}, \gamma_2^{(0)}, \gamma_3^{(0)} ]$ defines the initial point in the manifold and the initial condition $\vec{\alpha}^{(0)} = [ \alpha_1^{(0)}, \alpha_2^{(0)}, \alpha_3^{(0)} ]$ defines the initial tangent vector at the initial point (as the initial velocity vector). Making a simple analogy, the process can be understood as follows: suppose we place a tiny marble in the manifold defined by the parametric space of our random field model. At each iteration of the algorithm, the system of differential equations updates the position and the velocity vector of the tiny marble, based in the initial conditions. With these equations, it is possible to completely characterize the trajectory of the moving marble along geodesics in the underlying manifold (shortest paths). A limitation, however, is that, in order to compute the Christoffel symbols, we need to estimate the covariance matrix of the patches along the random field. In order to do so, we use Markov Chain Monte Carlo simulation to generate an outcome of the Gaussian random field at each iteration of the Runge-Kutta method using the parameter vector $\vec{\gamma}$, which is nothing more than the position of the marble in the manifold. Hence, as the tiny marble moves along the manifold, we have different outcomes for our random field model. At the end of the simulation, the arc length of the resulting geodesic curve defines our geodesic distance.

\begin{table}[ht]
\centering
\small
\caption{Results of the proposed iterative MCMC based fourth-order Runge-Kutta simulation for the Gaussian random fields manifold.}
\begin{tabular}{cccccccccccc}
\toprule
\multicolumn{3}{c}{\textbf{Init. point (A)}} & \multicolumn{3}{c}{\textbf{Initial} $\vec{\alpha}$} & \multicolumn{3}{c}{\textbf{Final point (B)}} & \multicolumn{3}{c}{\textbf{Distances}} \\
\midrule
$\mu$         & $\sigma^2$         & $\beta$  &  $\alpha_1$       & $\alpha_2$ & $\alpha_3$      & $\mu$  & $\sigma^2$      & $\beta$      & \textbf{Geo (AB)} & \textbf{Geo (BA)} & \textbf{Eucl.} \\
\midrule
0                & 1                    & 0                & 0.1                & 0.1               & 0.2                   & 0.7846               & 1.1079              & -0.9172                 & 1.7617                 &   1.8622   &  1.2126          \\
0                & 1                    & 0                & 0.01                & 0.02               & 0.08                   & 1.0057               & 0.9054              & -0.4039                 & 1.5918                 &   1.3889   &    1.0882        \\
0                & 1                    & 0                & 0.01                & 0.02               & 0.09                   & 1.1394               & 0.8927              & -0.4512                 & 1.7850                 &  2.2224    &    1.2305        \\
5.0                & 10.0                    & 0.5                & -0.1                & -0.1               & -0.2                   & 2.0308               & 12.7482              & 1.5256                 & 5.6583                 &  6.2141    &    4.1742        \\
10.0                & 5.0                    & 0.0                & 2.0                & 0.05               & 0.1                   & 10.9229               & 5.1097              & -1.3634                 & 2.1595                 &  1.8070    &    1.6397        \\
5.0                & 10.0                    & -0.5                & 0.01                & 0.01               & 0.2                   & 9.3335               & 17.7629              & -1.3973                 & 11.7472                 &   7.3078   &    8.9358        \\
5.0                & 10.0                    & -0.5                & 0.1                & 0.1               & 0.5                   & 7.6611               & 17.8300              & -2.2929                 & 10.1241                 &   10.844   &   8.4618        \\
10.0                & 100.0                    & 1.5                & -0.1                & -5.0               & -0.5                   & 5.3494               & 28.4040              & 1.8986                 & 71.759                 &   70.8599   &   70.0239        \\
0.0                & 1.0                    & 0.0                & 0.5                & 0.5               & 0.5                   & 0.7189               & 1.4497              & -1.6750                 & 2.4432                 &   2.3669   &   1.8788        \\
0.0                & 1.0                    & 0.0                & -1.0                & 0.25               & 1.0                   & -0.6844               & 0.9558              & -2.0375                 & 2.7383                 &   2.0959   &   2.1563        \\
0.0                & 1.0                    & 0.0                & 1.0                & 1.0               & 1.0                   & 0.7200               & 1.4367              & -2.3230                 & 3.1187                 &   2.4209   &   2.4757        \\
3.1415                & 9.8696                    & 0.0                & 0.01                & -0.02               & 0.08                   & 3.6003               & 11.5537              & 0.5078                 & 2.0014                 &  1.8976    &    1.8178        \\
\bottomrule
\end{tabular}
\label{tab:ds2}
\end{table}

\begin{figure}[ht]
\begin{center}
\includegraphics[scale=0.38]{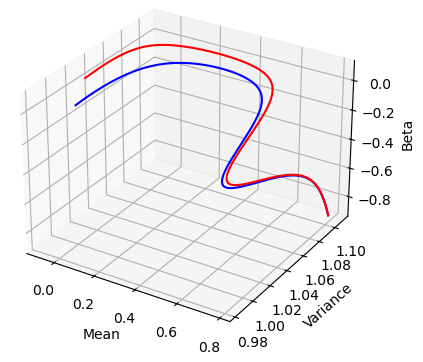}
\includegraphics[scale=0.25]{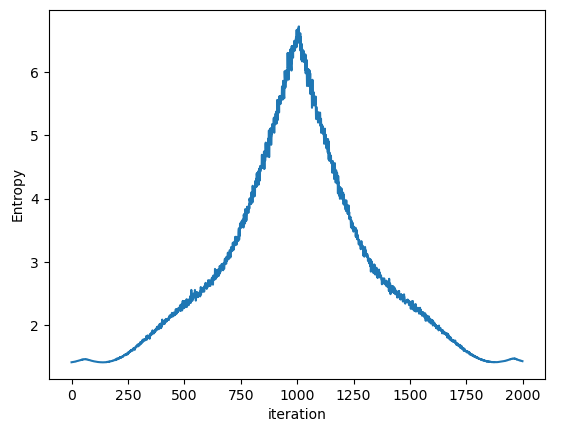}
\includegraphics[scale=0.25]{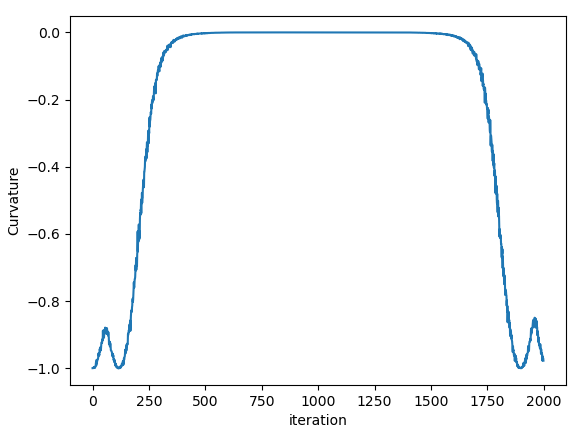}
\caption{Geodesic curves, entropy and curvature evolution in the Gaussian random fields manifold obtained by the proposed iterative MCMC based fourth-order Runge-Kutta algorithm and its time reverse simulations (first row in Table \ref{tab:ds2}). The blue curve denotes the orignal geodesic and the red curve denotes its time reversed version.}
\label{fig:C1}
\end{center}
\end{figure}

\begin{figure}[ht]
\begin{center}
\includegraphics[scale=0.38]{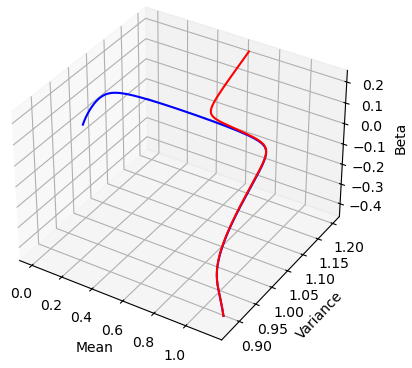}
\includegraphics[scale=0.25]{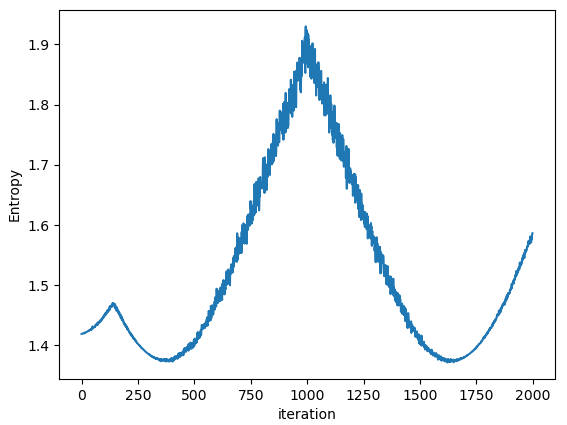}
\includegraphics[scale=0.25]{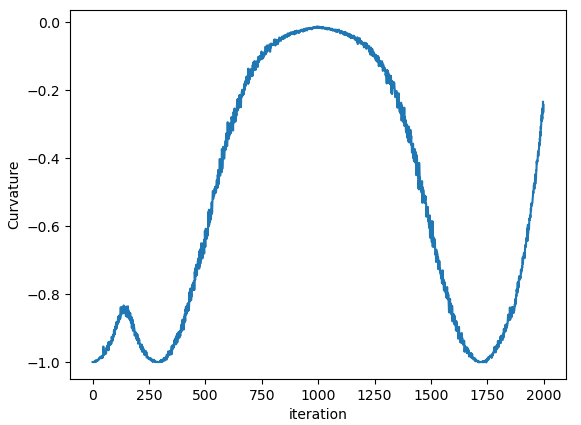}
\caption{Geodesic curves, entropy and curvature evolution in the Gaussian random fields manifold obtained by the proposed iterative MCMC based fourth-order Runge-Kutta algorithm and its time reverse simulations (third row in Table \ref{tab:ds2}). The blue curve denotes the orignal geodesic and the red curve denotes its time reversed version.}
\label{fig:C2}
\end{center}
\end{figure}

\begin{figure}[ht]
\begin{center}
\includegraphics[scale=0.37]{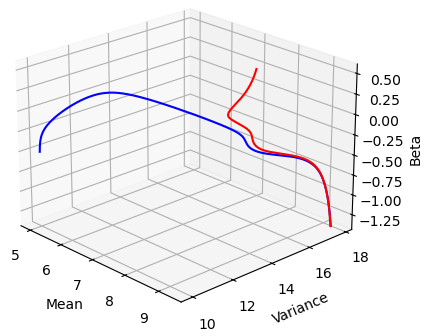}
\includegraphics[scale=0.25]{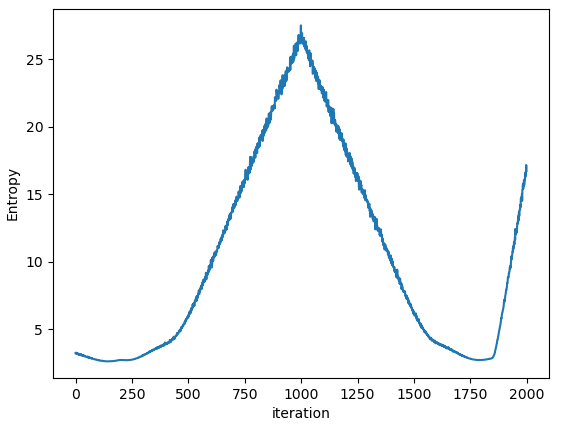}
\includegraphics[scale=0.25]{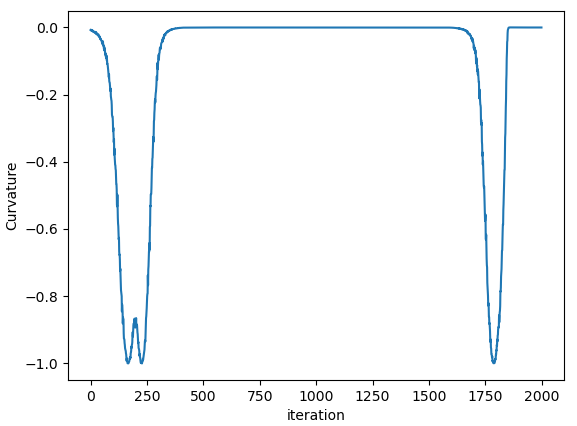}
\caption{Geodesic curves, entropy and curvature evolution in the Gaussian random fields manifold obtained by the proposed iterative MCMC based fourth-order Runge-Kutta algorithm and its time reverse simulations (sixth row in Table \ref{tab:ds2}). The blue curve denotes the orignal geodesic and the red curve denotes its time reversed version.}
\label{fig:C3}
\end{center}
\end{figure}

\begin{figure}[ht]
\begin{center}
\includegraphics[scale=0.38]{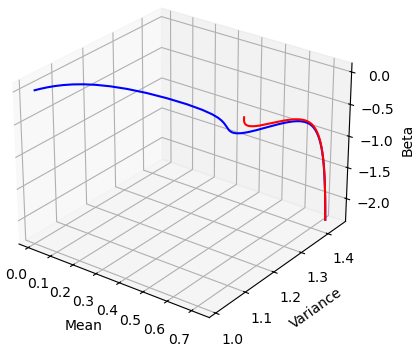}
\includegraphics[scale=0.25]{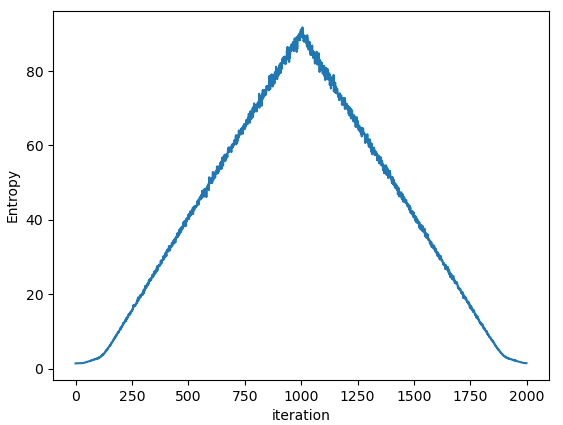}
\includegraphics[scale=0.25]{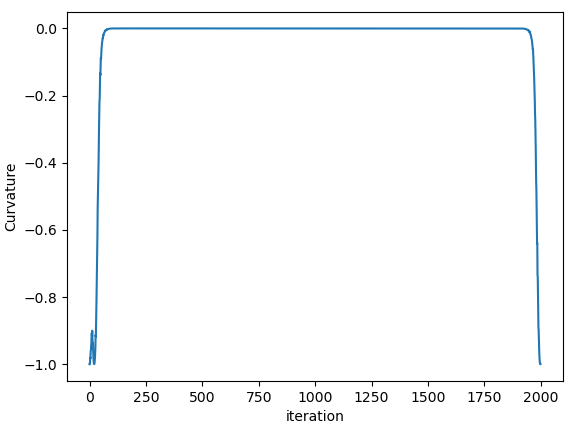}
\caption{Geodesic curves, entropy and curvature evolution in the Gaussian random fields manifold obtained by the proposed iterative MCMC based fourth-order Runge-Kutta algorithm and its time reverse simulations (eleventh row in Table \ref{tab:ds2}). The blue curve denotes the orignal geodesic and the red curve denotes its time reversed version.}
\label{fig:C4}
\end{center}
\end{figure}

Looking at the obtained results, it is possible to note a curious fact: the geodesic distances from the initial points A to the final points B are different from the reversed geodesic distances from final points B to initial points A, obtained by time reversing the Euler-Lagrange equations. In order to further investigate this behavior, we build the geodesic curves for the simulations described in Table \ref{tab:ds2}. Figures \ref{fig:C1}, \ref{fig:C2}, \ref{fig:C3} and \ref{fig:C4} show the geodesic curves, entropies and curvatures for some numerical simulations. Interestingly, a rather strange behavior is observed: the geodesic distances and their time reversed versions seem to diverge. We call this unexpected way of acting as the geodesic dispersion phenomenon in random fields dynamics. 

\begin{figure}[ht]
\begin{center}
\includegraphics[scale=0.35]{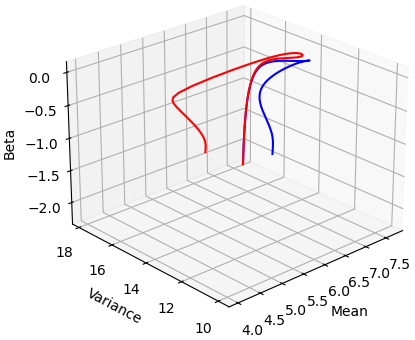}
\includegraphics[scale=0.35]{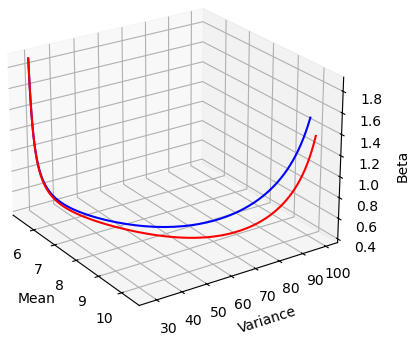}
\includegraphics[scale=0.35]{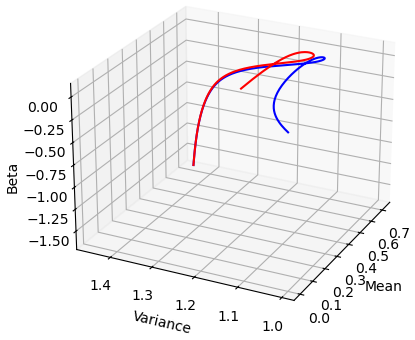}
\caption{Geodesic curves and their time reversed versions obtained by the proposed iterative MCMC based fourth-order Runge-Kutta algorithm regarding the rows seven, eight and nine of Table \ref{tab:ds2}) The blue curve denotes the orignal geodesic and the red curve denotes its time reversed version.}
\label{fig:GEO2}
\end{center}
\end{figure}

To further illustrate the variety of scenarios where this phenomenon occurs, Figure \ref{fig:GEO2} shows the geodesic curves and their time reversed versions for the simulations indicated by lines seven, eight and nine of Table \ref{tab:ds2}. In many situations, the dispersion of the time-reversed geodesic is highly significant, suggesting that the dynamics of the system is quite non-linear. To the best of our knowledge, this geodesic dispersion phenomenon has not been reported in the literature before. At the moment, there is no definitive explanation for the emergence of this phenomenon in random fields dynamics. Our conjecture is that it is related to the significant changes in the curvature of the parametric space along the trajectory of the geodesic curve described by the Euler-Lagrange equations. When the original geodesic curve is built by the numerical simulations and the system undergoes a phase transition, curvature goes from negative to zero, or vice-versa. The construction of the geodesic curve from A (initial point) to B (final point) starts in a region of the manifold that has negative curvature. It is known that negative curvature leads to hyperbolic geometry, in which two ``parallel'' geodesics tend to diverge. However, the construction of the time-reversed geodesic curve starts in a regions of the manifold that has zero curvature, in which two ``parallel'' geodesics remain parallel. When we move towards the region of negative curvature from the region of zero curvature, some degree of divergence is expected to happen during the process. Moreover, note that in Figures \ref{fig:C2}, \ref{fig:C3} and \ref{fig:C4}, it is possible to see some fluctuations in both entropy and curvature along the numerical simulations, leading to some asymmetries in the processes. For the interested reader, all the Python code used to generate the results presented in this paper, can be found at \url{https://github.com/alexandrelevada/GeodesicDispersion}.

\section{Conclusions and final remarks}

Random fields dynamics are present in many physical and complex systems due to their ability to capture the inherent variability and complexity in many natural phenomena as well as the stochastic nature of several physical process. Examples of random field applications in physics arise in a variety of fields, such as statistical physics, condensed matter physics, quantum field theory, chaos theory, high-energy physics, fluid dynamics, cosmology and astrophysics.

In this scientific investigation, our main motivating question was: can time irreversibility in Gaussian random fields dynamics be a consequence of the intrinsic geometry of their parametric space? By using MCMC simulation and the fourth-order Runke-Kutta method, we proposed an iterative algorithm to solve the Euler-Lagrange equations, after the derivation of the metric tensor and the Christoffel symbols of the metric. The analysis of the computational simulations revealed a strange fact: the geodesic dispersion phenomenon in random fields dynamics, which is related to how geodesic curves diverge from their time-reversed versions when the system undergoes phase transitions, due to abrupt changes in the local curvature of the underlying manifold. The obtained results from our numerical methods presented a series of intriguing patterns and behaviors that significantly contributed to our understanding of how random fields evolve in time and how irreversible dynamics can be a direct consequence of geometric properties. Through a meticulous exploration of the underlying mathematical framework, we have unveiled the intricate dynamics governing the geodesic dispersion, shedding light on its nuanced interplay with random fields.

Moreover, our research has established a formal connection between geodesic dispersion and the intrinsic geometric properties of the underlying space. By leveraging concepts from differential geometry, we have provided a mathematical foundation for the observed phenomena. This not only enhances our theoretical understanding but also opens avenues for developing more accurate models and predictive frameworks in the context of random fields. The implications of our findings may lead to relevant theoretical considerations in the study of several kinds of physical systems, elucidating how the passage of time in a system can be related to how the intrinsic geometry of its parametric space changes as we move through the underlying manifold. In random fields dynamics, the concepts of time and space seen to be intrinsically related.

In conclusion, our study on the geodesic dispersion phenomenon in random fields advances the frontier of knowledge in dynamic spatial processes. The integration of differential geometry, correlation analysis, and practical applications underscores the multidisciplinary significance of our findings. We hope that our work will stimulate further research, offering new perspectives and methodologies for investigating complex phenomena in physical systems.

Future works may include the derivation of the Ricci curvature tensor, which measures how a shape is deformed as one moves along geodesics in the space. Intead of using the simple Gaussian curvature, the more appropriate Ricci curvature is a more precise definition. The Ricci curvature of a manifold is often characterized as ``geodesic dispersion'', namely, whether two parallel geodesics shot from nearby points converge (spherical geometry or positive geometry), remain parallel (Euclidean geometry or zero curvature), or diverge (hyperbolic geometry or negative curvature). However, the components of the Ricci curvature tensor in Gaussian random fields depend on the derivatives of the Christoffel symbols, which are not simple to derive analytically.

\section*{Acknowledgments}
This study was partially funded by the Coordena\c{c}\~ao de Aperfei\c{c}oamento de Pessoal de N\'ivel Superior --- Brasil (CAPES), Finance Code 001.

\bibliographystyle{opticajnl}
\bibliography{sample}

\end{document}